\documentclass[conference,romanappendices,10pt]{IEEEtran}
\IEEEoverridecommandlockouts

\usepackage{algorithmic,amsmath,amssymb,amsthm,array,bbm,bibentry,cite,color,comment}
\usepackage{enumerate,enumitem,eurosym,float,graphicx,lettrine,mathrsfs,multirow,nomencl}
\usepackage{pict2e,psfrag,ragged2e,relsize,setspace,subfigure,supertabular,tabularx,url}
\usepackage[english]{babel}

{		\newtheorem{theorem}{Theorem}}
{ 		}
{ 		\newtheorem{lemma}{Lemma}}
{ 		}
{ 		\newtheorem{remark}{Remark}}
{		}
{		}
{ 		\newtheorem{corollary}{Corollary}}

\newcommand{\setL}{\mathcal{L}}

\newcommand{\Real}{\mbox{$\mathbb{R}$}}

\newcommand{\argmax}{\operatornamewithlimits{argmax}}

\newcommand{\diff}{\mathrm{d}}
\newcommand{\Exp}{\mathbb{E}}

\renewcommand{\Pr}{\mathbb{P}}

\newcommand{\sir}{\mathrm{SIR}}
\newcommand{\sinr}{\mathrm{SINR}}
\newcommand{\los}{\textnormal{\tiny{LOS}}}
\newcommand{\nlos}{\textnormal{\tiny{NLOS}}}
\newcommand{\Pcov}{\mathrm{P}_{\mathrm{cov}}}
\newcommand{\rmQ}{\textnormal{\tiny{Q}}}
\newcommand{\rmC}{\mathrm{C}}
\newcommand{\rmS}{\mathrm{S}}
\newcommand{\ASE}{\mathrm{ASE}}

\title{Performance Analysis of Ultra-Dense Networks with Elevated Base Stations}
\author{\IEEEauthorblockN{Italo Atzeni, Jes\'{u}s Arnau, and Marios Kountouris}
\IEEEauthorblockA{Mathematical and Algorithmic Sciences Lab \\ France Research Center, Huawei Technologies France SASU \\
Email: \{italo.atzeni, jesus.arnau, marios.kountouris\}@huawei.com}}

\begin{document}

\maketitle

\begin{abstract}
This paper analyzes the downlink performance of ultra-dense networks with elevated base stations (BSs). We consider a general dual-slope pathloss model with distance-dependent probability of line-of-sight (LOS) transmission between BSs and receivers. Specifically, we consider the scenario where each link may be obstructed by randomly placed buildings. Using tools from stochastic geometry, we show that both coverage probability and area spectral efficiency decay to zero as the BS density grows large. Interestingly, we show that the BS height alone has a detrimental effect on the system performance even when the standard single-slope pathloss model is adopted.
\end{abstract}

\vspace{1mm}


\vspace{-2mm}

%=========================================================================
\section{Introduction} \label{sec:intro}
%=========================================================================

Recently, there has been an increasing interest in network densification as a means to fulfill the performance requirements of the 5th generation (5G) of wireless networks \cite{Bhu14}. In particular, ultra-dense networks (UDNs), i.e., dense and massive deployments of base stations (BSs) and access points (APs), are regarded as key enablers to provide higher data rates and enhanced coverage by exploiting spatial reuse while retaining, at the same time, seamless connectivity and low energy consumption.

In parallel with the aforementioned interest for UDNs, mainly motivated by recent theoretical results \cite{And11, Dhi12}, many efforts have recently been carried out to analyze the effect of densifying the network under realistic propagation models. In this respect, \cite{Zha15} studies the impact of dual-slope pathloss on the performance of downlink UDNs assuming a critical distance after which transmissions become non-line-of-sight (NLOS), and shows that both coverage and capacity strongly depend on the network density. More involved models can be found, e.g., in \cite{Gal15,Din16,Gup16}, where the pathloss exponent changes with a probability that depends on the distance between BSs and user equipments (UEs); in addition to modifying the pathloss exponent with such distance-dependent probability, \cite{Arn16,Atz17b,And16} also consider varying the type of fading. Coverage and rate scaling laws in UDNs are derived in \cite{Ngu16}.

In this work, we incorporate the BS height into the discussion on whether densifying the network will provide monotonically increasing data rates or not. We start from a general pathloss model with distance-dependent probability of line-of-sight (LOS) transmission. More specifically, we consider the practically relevant scenario where each link may be obstructed by randomly placed buildings, i.e., LOS transmission occurs only when no building cuts the line between the elevated BS and the UE.\footnote{For the blockage effect of buildings at high frequencies, see \cite{Bai14}.} Consistently with previous works, our derivations show that the area spectral efficiency (ASE) does not increase monotonically with the BS density. Moreover, we reveal something even more surprising: the BS height alone has a detrimental effect on the system performance even when the standard single-slope pathloss model is adopted, causing the ASE to decay to zero as the BS density grows large; this corroborates the findings in \cite{Din16a}, appeared during the preparation of this manuscript. It follows that, with elevated BSs, densifying the network eventually leads to near-universal outage. In other words, BSs should be mounted at the same height as the UEs for higher coverage and throughput gains \cite{Atz17b}.

\begin{figure*}
\addtocounter{equation}{+5}
\begin{align}
\label{eq:P_cov} \Pcov (\theta) & = \int_{0}^{\infty} \bigg( p_{\los}(r) \setL_{I} \bigg( \frac{\theta}{\ell_{\los}(r,h)} \bigg) + \big( 1 - p_{\los}(r) \big) \setL_{I} \bigg( \frac{\theta}{\ell_{\nlos}(r,h)} \bigg) \bigg) \phi(r) \diff r \\
\label{eq:L_I} \setL_{I} (s) & \triangleq \setL_{I}^{\nlos} (s) \exp \bigg( - 2 \pi \lambda \int_{\nu(r)}^{\infty} p_{\los}(t) \bigg( \frac{1}{1 + s \ell_{\nlos}(t,h)} - \frac{1}{1 + s \ell_{\los}(t,h)} \bigg) t \diff t \bigg)
\end{align}
\addtocounter{equation}{-7} \vspace{-1mm}
\hrulefill \vspace{-1mm}
\end{figure*}

%=========================================================================
\section{System Model} \label{sec:syst}
%=========================================================================

We consider a typical downlink UE located at the origin of the Euclidean plane. The location distribution of the BSs is modeled using the marked Poisson point process (PPP) $\widehat{\Phi} \triangleq \{(x_{i}, g_{x_{i}})\} \subset \Real^{2} \times \Real^{+}$, where the underlying point process $\Phi \triangleq \{ x_{i} \} \subset \Real^{2}$ is a homogeneous PPP with density $\lambda$, measured in [BS/m$^{2}$], and the mark $g_{x_{i}} \in \Real^{+}$ represents the channel power fading gain from the BS located at $x_{i}$ to the typical UE. We assume that all BSs are at the same height $h \geq 0$, measured in [m], whereas the typical UE is at the ground level; alternatively, $h$ can be interpreted as the elevation difference between BSs and UEs if the latter are all at the same height. Furthermore, we assume that all channel amplitudes are Rayleigh distributed so that $g_{x_{i}} \sim \exp(1)$, $\forall x_{i} \in \Phi$; the~longer version of this paper \cite{Atz17b} also considers Nakagami-$m$ fading.

Let $r_{x} \triangleq \| x \|$ denote the horizontal distance of $x$ from the typical UE, measured in [m]. Assuming the standard power-law pathloss model, we have the following pathloss functions: $\ell_{\los}(r_{x},h) \triangleq (r_{x}^2 + h^2)^{-\frac{\alpha_{\los}}{2}}$ if $x$ is in LOS conditions and $\ell_{\nlos}(r_{x},h) \triangleq (r_{x}^2 + h^2)^{-\frac{\alpha_{\nlos}}{2}}$ if $x$ is in NLOS conditions, with $\alpha_{\nlos} > \alpha_{\los} > 2$. We consider a distance-dependent LOS probability function $p_{\los}(r_{x})$, i.e., the probability that a BS located at $x$ experiences LOS propagation depends on the distance $r_{x}$.\footnote{Observe that the dual-slope model presented in \cite{Zha15} is a particular case of the proposed framework with $h=0$ and $p_{\los}(r_{x} \leq R_{\mathrm{c}}) = 1$ and $p_{\los}(r_{x} > R_{\mathrm{c}}) = 0$ for a certain critical distance $R_{\mathrm{c}}$.} We remark that each BS is characterized by either LOS or NLOS propagation independently from the others and regardless of its role as serving or interfering BS.

Let $\Phi_{\los} \triangleq \{ x  \in \Phi : x {\rm \ in \ LOS} \}$ and let $\Phi_{\nlos} \triangleq \Phi \, \backslash \, \Phi_{\los}$. The signal-to-interference-plus-noise ratio (SINR) when the typical UE is associated to the BS located at $x$ is given by
\begin{align} \label{eq:SINR}
\sinr_{x} \triangleq \frac{g_{x} \ell_{\rmQ}(r_{x},h)}{I + \sigma^{2}}
\end{align}
where the sub-index $\mathrm{Q}$ takes the form $\mathrm{Q} = \mathrm{LOS}$ if $x \in \Phi_{\los}$ and $\mathrm{Q} = \mathrm{NLOS}$ if $x \in \Phi_{\nlos}$, $I$ is the aggregate interference term defined as
\begin{align} \label{eq:I}
I \triangleq \sum_{y \in \Phi_{\los} \backslash \{x\}} g_{y} \ell_{\los}(r_{y},h) + \sum_{y \in \Phi_{\nlos} \backslash \{x\}} g_{y} \ell_{\nlos}(r_{y},h)
\end{align}
and $\sigma^{2}$ is the additive noise power. For simplicity, we consider the interference-limited case, i.e., $I \gg \sigma^{2}$, and we thus focus on the signal-to-interference ratio (SIR). Our analysis can be extended with more involved calculations to the general case.

%=========================================================================
\section{Coverage Probability} \label{sec:cov}
%=========================================================================

In this section, we formalize the coverage probability when serving and interfering BSs independently experience LOS or NLOS conditions with respect to the typical UE depending on their distance from the latter. The coverage probability is defined as the probability that the received SIR is larger than a target SIR threshold $\theta$, i.e., $\Pcov (\theta) \triangleq \Pr [\sir_{x} > \theta]$.

We consider a unified framework that encompasses closest and strongest (i.e., highest SINR) BS association. For this purpose, we introduce the following preliminary definitions~\cite{Arn16,Atz17b}: \vspace{-1mm}
\begin{align}
\label{eq:phi} \phi(r) & \triangleq \left\{
\begin{array}{ll}
2 \pi \lambda e^{- \pi \lambda r^{2}} r, & \quad \textrm{closest BS} \\
2 \pi \lambda r, & \quad \textrm{strongest BS}
\end{array} \right. \\
\label{eq:nu} \nu(r) & \triangleq \left\{
\begin{array}{ll}
r, & \hspace{16mm} \quad \textrm{closest BS} \\
0, & \hspace{16mm} \quad \textrm{strongest BS}.
\end{array} \right.
\end{align}
Note that \eqref{eq:phi} for closest BS association gives the pdf of the distance between the typical UE and the serving BS~\cite{And11}, which is not the case for strongest BS association \cite{Dhi12}. In addition, let $\setL_{I}^{\rmQ} (s)$ denote the Laplace transform of the interference when all the interfering BSs are in either LOS or NLOS conditions:
\begin{align} \label{eq:L_I_Q}
\setL_{I}^{\rmQ} (s) \triangleq \exp \bigg( - 2 \pi \lambda \int_{\nu(r)}^{\infty} \bigg( 1 - \frac{1}{1 + s \ell_{\rmQ} (t,h)} \bigg) t \diff t \bigg).
\end{align}

%=========================================================================
\subsection{General LOS/NLOS Model} \label{sec:cov_general}
%=========================================================================

We begin by considering a general expression of $p_\los(r)$. The resulting coverage probability is given in the next theorem.

\begin{theorem} \label{th:P_cov} \rm{
The coverage probability and the Laplace transform of the interference are given by \eqref{eq:P_cov} and \eqref{eq:L_I}, respectively, at the top of the page, with $\setL_{I}^{\nlos} (s)$ defined in \eqref{eq:L_I_Q}.}
\end{theorem}
\addtocounter{equation}{+2}

\begin{IEEEproof}
See Appendix~\ref{sec:A1}.
\end{IEEEproof} \vspace{1mm}

\begin{remark} \rm{
Since $\alpha_{\nlos} > \alpha_{\los}$, the argument of the exponential function in \eqref{eq:L_I} is always negative and, in consequence, $\setL_{I} (s) \leq \setL_{I}^{\nlos} (s)$; this is due to the contribution from the interfering BSs in LOS conditions. On the other hand, the possibility of LOS desired signal enhances the coverage probability in \eqref{eq:P_cov}.}
\end{remark}

It is not straightforward to get insights on which of the two aforementioned effects is dominant. In Section~\ref{sec:num}, we will show some trends through numerical evaluation of the aforementioned expressions using the model described next.

%=========================================================================
\subsection{LOS/NLOS Model with Randomly Placed Buildings} \label{sec:cov_buildings}
%=========================================================================

So far we have assumed no particular expression for $p_\los(r)$. In this section we introduce a model for this quantity that takes into account the combined influence of the link distance and the BS height through the probability of the link being blocked by a building.\footnote{Other options exist in the literature: for instance, in \cite{And16a}, the BS height, the link distance, and the pathloss exponent are related through the effect of the ground-reflected ray.}

\begin{figure}[t!]
\centering
\includegraphics[scale=0.95]{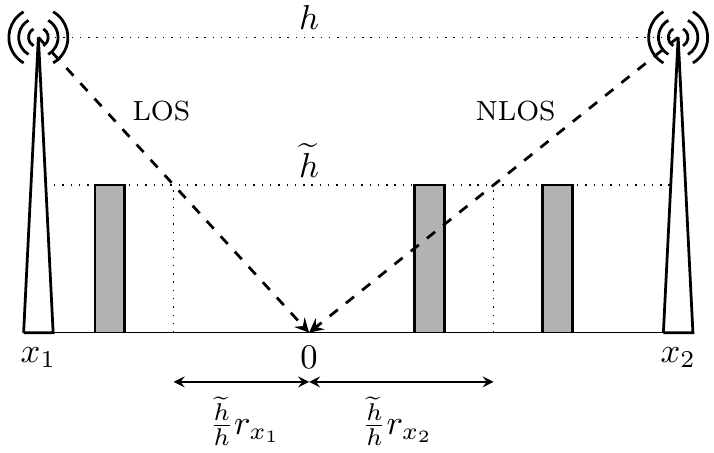}
\caption{LOS/NLOS model with buildings randomly placed between the BSs and the typical UE.} \label{fig:height} \vspace{-2mm}
\end{figure}

Given a point $x$, we assume that buildings with fixed height $\widetilde{h}$ are randomly placed between $x$ and the typical UE. If the straight line between the elevated BS at $x$ and the typical UE does not cross any buildings, then the transmission occurs in LOS conditions; alternatively, if at least one building cuts this straight line, then the transmission occurs in NLOS conditions.\footnote{The cumulative effect of multiple obstacles is considered in \cite{Lee16}.} A simplified example is illustrated in Figure~\ref{fig:height}. Note that, in this context, the probability of $x$ being in LOS conditions depends not only on the distance $r_{x}$, but also on the parameter $\tau \triangleq \min \big( \frac{\widetilde{h}}{h}, 1 \big)$. More precisely, the LOS probability corresponds to the probability of having no buildings in the segment of length $\tau r_{x}$ next to the typical UE.

If the location distribution of the buildings follows a one-dimensional PPP with density $\widetilde{\lambda}$, measured in [buildings/m], the LOS probability function is given by $p_{\los}(r_{x}, \tau) = e^{-\widetilde{\lambda} \tau r_{x}}$. Observe that $p_{\los}(r_{x}, \tau) = 1$ (all links are in LOS) when $\widetilde{\lambda} = 0$ or $\widetilde{h} = 0$, whereas $p_{\los}(r_{x}, \tau) = 0$ (all links are in NLOS) when $\widetilde{\lambda} \to \infty$. In Section~\ref{sec:num}, we will numerically illustrate the effect of different building densities and comment on the interplay between LOS/NLOS desired signal and interference.

%=========================================================================
\section{The Effect of BS Height} \label{sec:height}
%=========================================================================

In this section, we study the effect of BS height alone. For this purpose, we consider the single-slope pathloss function $\ell (r_x, h) \triangleq (r_{x}^2 + h^2)^{-\frac{\alpha}{2}}$.

%=========================================================================
\subsection{Impact on Interference} \label{sec:height_int}
%=========================================================================

The interference with elevated BSs is characterized here. In doing so, we analyze closest and strongest BS association separately by introducing the super-indices $\rmC$ and $\rmS$, respectively. We begin by observing that $\ell (r_{x}, h)$ yields a bounded pathloss model for any BS height $h > 0$, since the BSs cannot get closer than $h$ to the typical UE (this occurs when $r_{x} = 0$). 

Recall that the notation $\nu(r)$ encompasses both closest and strongest BS association, and let $\setL_{I,0} (s)$ denote the Laplace transform of the interference with $h=0$, i.e.,
\begin{align} \label{eq:L_I_0}
\setL_{I,0} (s) \triangleq \exp \bigg( - 2 \pi \lambda \int_{\nu(r)}^{\infty} \bigg( 1 - \frac{1}{1 + s t^{-\alpha}} \bigg) t \diff t \bigg).
\end{align}
The following lemma expresses the Laplace transforms of the interference with any $h$.
\begin{lemma} \label{lem:Laplace} {\rm
For closest and strongest BS association, the Laplace transforms of the interference with elevated BSs can be written as
\begin{align}
\label{eq:L_I_0_C} \setL_{I}^{(\rmC)} (s) & \triangleq \setL_{I,0}^{(\rmC)}(s) \exp \bigg( 2 \pi \lambda \int_{r}^{\sqrt{r^{2}+h^{2}}} \! \! \bigg( 1 - \frac{1}{1 + s t^{-\alpha}} \bigg) t \diff t \bigg) \\
\label{eq:L_I_0_S} \setL_{I}^{(\rmS)} (s) & \triangleq \setL_{I,0}^{(\rmS)}(s) \exp \bigg( 2 \pi \lambda \int_{0}^{h} \bigg( 1 - \frac{1}{1 + s t^{-\alpha}} \bigg) t \diff t \bigg)
\end{align}
respectively, with $\setL_{I,0}^{(\rmC)}$ and $\setL_{I,0}^{(\rmS)}$ defined in \eqref{eq:L_I_0}.}
\end{lemma}

\begin{IEEEproof}
The Laplace transforms of the interference \eqref{eq:L_I_0_C} and \eqref{eq:L_I_0_S} can be obtained from
\begin{align}
\setL_{I} (s) \triangleq \exp \bigg( - 2 \pi \lambda \int_{\nu(r)}^{\infty} \bigg( 1 - \frac{1}{1 + s \ell(t,h)} \bigg) t \diff t \bigg)
\end{align}
first by substituting $\sqrt{t^{2} + h^{2}} \to q$ and then by splitting the integration intervals in two parts.
\end{IEEEproof} \vspace{1mm}

From \eqref{eq:L_I_0_C} and \eqref{eq:L_I_0_S}, it is straightforward to see that the interference is reduced when $h>0$ with respect to when $h=0$, since the original Laplace transforms are multiplied by exponential terms with positive arguments. 

For strongest BS, we provide a further interesting result on the expected interference power. Recall that, for strongest BS association and for $h=0$, the expected interference power is infinite \cite[Ch.~5.1]{Hae12}. Let $U(a,b,z) \triangleq \frac{1}{\Gamma(a)} \int_{0}^{\infty} e^{-z t} t^{a - 1} (1 + t)^{b - a - 1} \diff t$ denote Tricomi's confluent hypergeometric function and let $E_{n}(z) \triangleq \int_{1}^{\infty} e^{-z t} t^{-n} \diff t$ be the exponential integral function. A consequence of the bounded pathloss model is given in the following lemma, which characterizes the expected interference with elevated BSs for strongest BS association.

\begin{lemma} \label{lem:interference} \rm{
For strongest BS association, the expected interference power with elevated BSs is finite and is given by
\begin{align} \label{eq:interference}
\Exp \bigg[ \sum_{y \in \Phi \backslash \{x\}} \! \! g_{y} \ell(r_{y},h) \bigg] \! \! < \! \sum_{i=1}^{\infty} (\pi \lambda)^{i} h^{2 i - \alpha} U \big( i, i + 1 - \tfrac{\alpha}{2}, \pi \lambda h^{2} \big)
\end{align}
where the expected interference power from the closest interfering BS, located at $x_{1}$, corresponds to
\begin{align} \label{eq:interference_nearest}
\Exp \big[ g_{x_{1}} \ell(r_{y_{1}},h) \big] = \pi \lambda h^{2 - \alpha} e^{\pi \lambda h^{2}} E_{\frac{\alpha}{2}}(\pi \lambda h^{2}).
\end{align}}
\end{lemma}

\begin{IEEEproof}
See Appendix~\ref{sec:A2_interference}.
\end{IEEEproof} \vspace{1mm}

%=========================================================================
\subsection{Impact on Coverage Probability and ASE} \label{sec:height_cov}
%=========================================================================

We now focus on the effect of the BS height on the coverage probability. We use $\mathrm{P}_{\mathrm{cov},0}^{(\rmC)} (\theta)$ and $\mathrm{P}_{\mathrm{cov},0}^{(\rmS)} (\theta)$ to denote the coverage probabilities for closest and strongest BS association when $h=0$: these are known to be independent on $\lambda$ and can be written as (see \cite{And11,Dhi12}, respectively, for details)
\begin{align}
\label{eq:P_cov_C_0} \mathrm{P}_{\mathrm{cov},0}^{(\rmC)} (\theta) & \triangleq \frac{1}{\psi(\theta) + 1} \\
\label{eq:P_cov_S_0} \mathrm{P}_{\mathrm{cov},0}^{(\rmS)} (\theta) & \triangleq \frac{\alpha \sin \big( \frac{2 \pi}{\alpha} \big)}{2 \pi \theta^{\frac{2}{\alpha}}}
\end{align}
with
\begin{align} \label{eq:psi}
\psi(z) \triangleq \frac{2 z}{\alpha-2} \;_{2}F_{1} \big( 1, 1 - \tfrac{2}{\alpha}, 2 - \tfrac{2}{\alpha}, - z \big)
\end{align}
where $_{2}F_{1} (a,b,c,z)$ is the Gauss hypergeometric function. Furthermore, it is convenient to examine the achievable ASE as an additional performance metric and we thus define $\ASE (\theta, \lambda) \triangleq \lambda \Pcov (\theta, \lambda) \log_{2} (1 + \theta)$, measured in [bps/Hz/m$^{2}$].

For the case of closest BS association, Theorem~\ref{th:height} and Corollary~\ref{cor:lambda_opt} provide, respectively, closed-form and integral expressions for the coverage probability and for the optimal BS density in terms of ASE with any $h$.

\begin{figure*}[t!]
\centering
\includegraphics[scale=0.95]{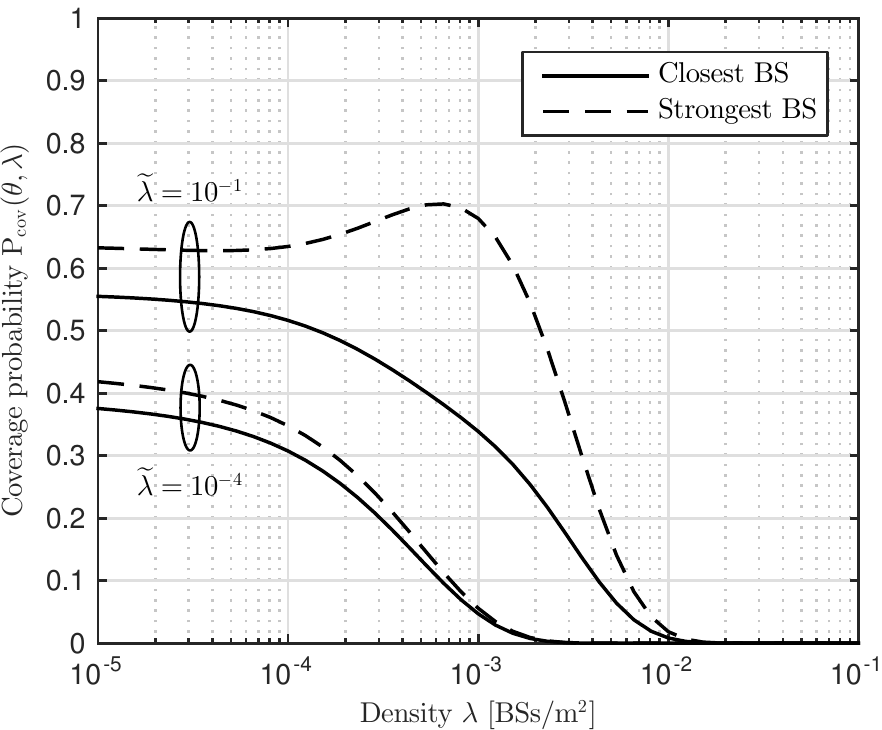} \hspace{4mm}
\includegraphics[scale=0.95]{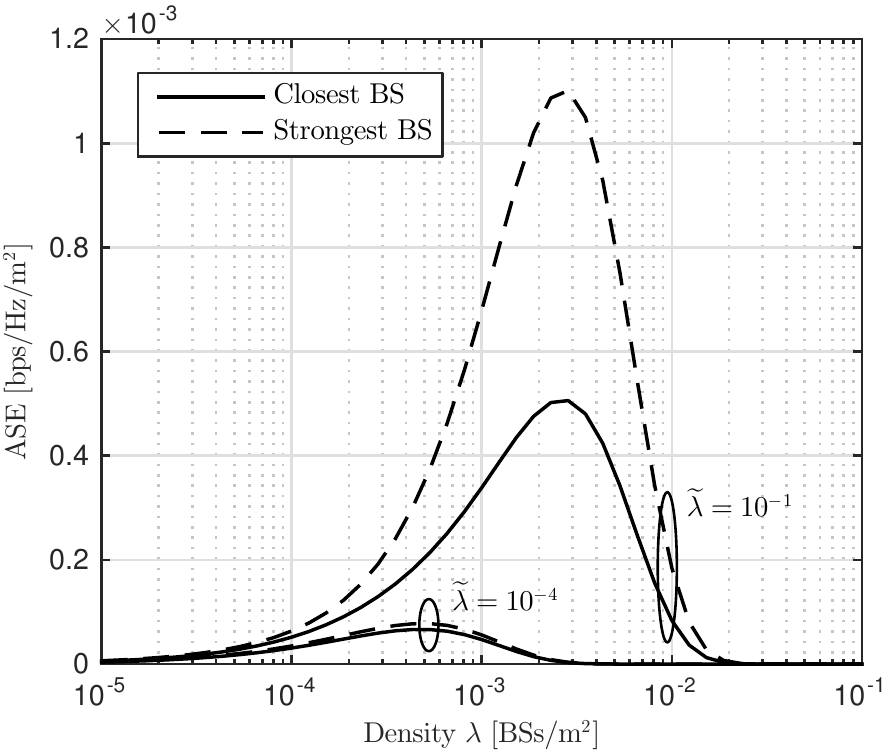}
\caption{Coverage probability (left) and ASE (right) versus BS density $\lambda$ with $h=20$~m and $\widetilde{h}=10$~m, for $\widetilde{\lambda} \! = \! 10^{-4}$~buildings/m and $\widetilde{\lambda} \! = \! 10^{-1}$~buildings/m.} \label{fig:num_model} \vspace{-2mm}
\end{figure*}

\begin{theorem} \label{th:height} \rm{
For closest and strongest BS association, the coverage probabilities with elevated BSs are given by
\begin{align}
\label{eq:P_cov_C} \Pcov^{(\rmC)} (\theta, \lambda) & = \mathrm{P}_{\mathrm{cov},0}^{(\rmC)} (\theta) \exp \big( - \pi \lambda h^{2} \psi(\theta) \big) \\
\label{eq:P_cov_S} \Pcov^{(\rmS)} (\theta, \lambda) & = 2 \pi \lambda \int_{h}^{\infty} \exp \big( - \pi \lambda h^{2} \psi(\theta h^{-\alpha} r^{\alpha}) \big) r \diff r
\end{align}
respectively, with $\mathrm{P}_{\mathrm{cov},0}^{(\rmC)} (\theta)$ and $\psi(z)$ defined in \eqref{eq:P_cov_C_0} and \eqref{eq:psi}, respectively.}
\end{theorem}

\begin{IEEEproof}
See Appendix~\ref{sec:A2_height}.
\end{IEEEproof} \vspace{1mm}

\begin{corollary} \label{cor:lambda_opt} \rm{
Let $\ASE^{(\rmC)}(\theta, \lambda) \triangleq \lambda \Pcov^{(\rmC)} (\theta, \lambda) \log_{2} (1 + \theta)$ denote the achievable ASE for closest BS association. Then, we have
\begin{align}
\lambda_{\mathrm{opt}}^{(\rmC)} \triangleq \argmax_{\lambda} \ASE^{(\rmC)}(\theta, \lambda) = \frac{1}{\pi h^{2} \psi(\theta)}.
\end{align}
}
\end{corollary}

\begin{IEEEproof}
The optimal BS density $\lambda_{\mathrm{opt}}^{(\rmC)}$ can be easily obtained as the solution of $\frac{\diff}{\diff \lambda} \lambda \Pcov^{(\rmC)} (\theta, \lambda) = 0$.
\end{IEEEproof} \vspace{1mm}

Theorem~\ref{th:height} unveils the detrimental effect of BS height on the system performance. This degradation stems from the fact that the BS height affects more the distance of the typical UE from its serving BS than the distances from the interfering BSs: hence, desired signal power and interference power do not grow at the same rate as when $h = 0$. The following lemma strengthens this claim by showing the asymptotic performance for both closest and strongest BS association.

\begin{lemma} \label{lem:limits} \rm{
For any BS height $h > 0$, the following holds:
\begin{itemize}
\item[(i)] $\lim_{\lambda \to 0} \mathrm{P}_{\mathrm{cov}}^{(\rmC)} (\theta, \lambda) = \mathrm{P}_{\mathrm{cov},0}^{(\rmC)} (\theta)$;
\item[(ii)] $\lim_{\lambda \to 0} \mathrm{P}_{\mathrm{cov}}^{(\rmS)} (\theta, \lambda) = \mathrm{P}_{\mathrm{cov},0}^{(\rmS)} (\theta)$;
\item[(iii)] $\lim_{\lambda \to \infty} \mathrm{P}_{\mathrm{cov}}^{(\rmC)} (\theta, \lambda) = \lim_{\lambda \to \infty} \mathrm{P}_{\mathrm{cov}}^{(\rmS)} (\theta,\lambda) = 0$.
\end{itemize}
}
\end{lemma}

\begin{IEEEproof}
See Appendix~\ref{sec:A2_limits}.
\end{IEEEproof} \vspace{1mm}

\begin{remark} \rm{
For a fixed BS height $h > 0$, the coverage probability monotonically decreases as the BS density $\lambda$ increases: consequently, the ASE goes to zero as $\lambda \to \infty$. On the other hand, the effect of BS height becomes negligible as $\lambda \to 0$.}
\end{remark}

\noindent In practice, we will see in Section~\ref{sec:num} that the coverage probability and the ASE decays to zero even for moderately low BS densities (i.e., for $\lambda \sim 10^{-3}$~BSs/m$^2$).

\begin{remark} \rm{
For a fixed BS density $\lambda$, the coverage probability monotonically decreases as the BS height $h$ increases. More specifically, Corollary~\ref{cor:lambda_opt} implies that $\ASE^{(\rmC)}(\theta,\lambda_{\mathrm{opt}}^{(\rmC)}) \propto \frac{1}{h^{2}}$. Therefore, the optimal BS height is~$h=0$.}
\end{remark}

\noindent When serving and interfering BSs are characterized by the same LOS/NLOS conditions or, more generally, by the same distance-dependent LOS probability function (as the one described in Section~\ref{sec:cov_buildings}), the optimal BS height is always $h = 0$, confirming the findings of \cite{Din16a}. However, under a propagation model where the interfering BSs are always in NLOS conditions and the serving BS can be either in LOS or NLOS conditions, a non-zero optimal BS height is expected: in fact, in this case, there would be a tradeoff between pathloss (for which a low BS is desirable) and probability of LOS desired signal (for which a high BS is desirable) \cite{Atz17b}.

%=========================================================================
\section{Numerical Results} \label{sec:num}
%=========================================================================

In this section, we consider the LOS/NLOS model with randomly placed buildings proposed in Section~\ref{sec:cov_buildings} and evaluate the coverage probability and the ASE. We consider the following parameters: the pathloss exponents are fixed to $\alpha_{\los} = 3$ and $\alpha_{\nlos} = 4$, whereas the SIR threshold is $\theta =1$.

Figure~\ref{fig:num_model} plots the coverage probability and the ASE versus the BS density $\lambda$ with BS height $h=20$~m and building height $\widetilde{h}=10$~m; two building densities are considered, i.e., $\widetilde{\lambda} = 10^{-4}$~buildings/m and $\widetilde{\lambda} = 10^{-1}$~buildings/m. It is straightforward to note that a high building density is beneficial in this setting since it creates nearly NLOS conditions, whereas a low building density resembles a nearly LOS scenario. Moreover, for $\widetilde{\lambda} = 10^{-1}$~buildings/m and strongest BS association, we observe a peak in the coverage probability around $\lambda = 10^{-3}$~BSs/m$^{2}$ due to the choice of the parameters $h$ and $\widetilde{h}$.

\begin{figure}[t!]
\centering
\includegraphics[scale=0.95]{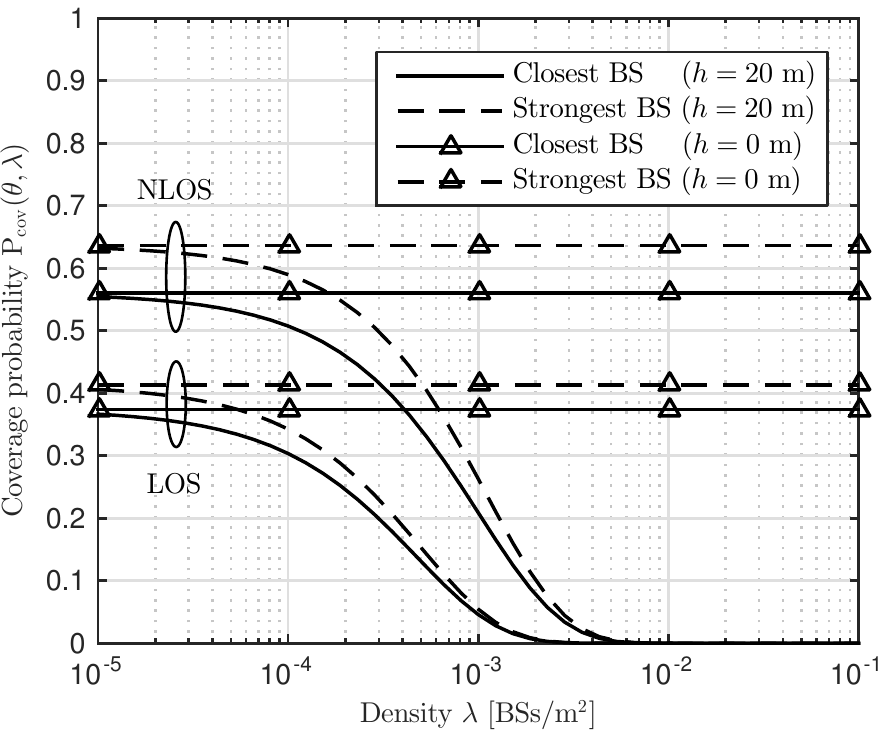}
\caption{Effect of BS height with LOS and NLOS: coverage probability versus BS density $\lambda$, for $h=20$~m and $h=0$~m.} \label{fig:num_height1} \vspace{-2mm}
\end{figure}

\begin{figure}[t!]
\centering \vspace{-2.2mm}
\includegraphics[scale=0.95]{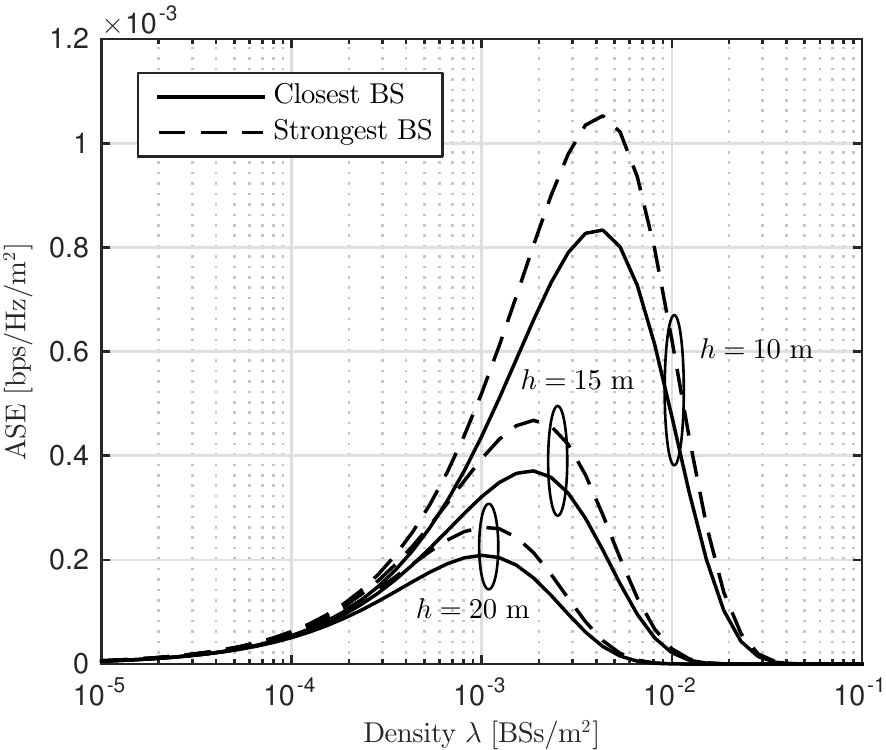}
\caption{Effect of BS height with NLOS: ASE versus BS density $\lambda$, for $h=10$~m, $h=15$~m, and $h=20$~m.} \label{fig:num_height2} \vspace{-3.5mm}
\end{figure}

We now focus on the effect of the BS height alone. In the following, we refer to the LOS (resp. NLOS) case where both serving and interfering BSs are in LOS (resp. NLOS) conditions. Figure~\ref{fig:num_height1} considers $h=20$~m and shows that the coverage probability goes to zero even for moderately low BS densities, i.e., at $\lambda \simeq 2 \times 10^{-3}$~BSs/m for the LOS case and at $\lambda \simeq 5 \times 10^{-3}$~BSs/m for the NLOS case. For comparison, the coverage probabilities with $h=0$~m are also plotted (see $\mathrm{P}_{\mathrm{cov},0}^{(\rmC)} (\theta)$ and $\mathrm{P}_{\mathrm{cov},0}^{(\rmS)} (\theta)$ in \eqref{eq:P_cov_C_0} and \eqref{eq:P_cov_S_0}, respectively): as expected, the coverage probability with elevated BSs approaches that with $h=0$ as $\lambda \to 0$. In Figure~\ref{fig:num_height2}, we show the ASE for three different values of the BS height, namely: $h = 10$~m, $h = 15$~m, and $h = 20$~m. Interestingly, the BS density that maximizes the ASE for strongest BS association coincides with the optimal density for the case of closest BS association, where the latter can be computed in closed form as in Corollary~\ref{cor:lambda_opt}.

%=========================================================================
\section{Conclusions} \label{sec:conc}
%=========================================================================

In this paper, we study the downlink performance of UDNs with elevated BSs. We assume a general dual-slope pathloss model with distance-dependent probability of LOS transmission between BSs and UEs and we obtain expressions for the coverage probability and ASE using tools from stochastic geometry. In particular, we consider the scenario where each link may be obstructed by randomly placed buildings. The most important implication is that, with elevated BSs, densifying the network eventually leads to near-universal outage.

In the settings of this paper the optimal BS height turned out to be always zero, i.e., at the same level as the UE height. However, under a propagation model where the interfering BSs are always in NLOS conditions and the serving BS can be either in LOS or NLOS conditions, a non-zero optimal BS height is expected as a result of the tradeoff between pathloss and probability of LOS desired signal.

\appendices

%=========================================================================
\section{Proof of Theorem~\ref{th:P_cov}} \label{sec:A1}
%=========================================================================

The coverage probability is given by
\begin{align}
\Pcov (\theta) & = \Pr [\sir_{x} > \theta] \\
%& = \Pr \bigg[ g_{x} > \frac{\theta I}{\ell_{\rmQ}(r_{x},h)} \bigg] \\
& = \int_{0}^{\infty} \Pr \bigg[ g_{x} > \frac{\theta I}{\ell_{\rmQ}(r,h)} \Big| r \bigg] \phi(r) \diff r 
\end{align}
and, since $\mathrm{Q} = \mathrm{LOS}$ with probability $p_\los(r)$ and $\mathrm{Q} = \mathrm{NLOS}$ with probability $1 - p_\los(r)$, the expression in \eqref{eq:P_cov} readily follows. On the other hand, the Laplace transform in \eqref{eq:L_I} is obtained through the following steps:
\begin{align}
\setL_{I} (s) & = \Exp [e^{-s I}] \\
\nonumber & = \Exp_{\Phi} \bigg[ \prod_{y \in \Phi_{\los} \backslash \{x\}} \Exp_{g_{y}} \big[ \exp \big( - s g_{y} \ell_{\los}(r_{y},h) \big) \big] \\
& \phantom{=} \ \times \prod_{y \in \Phi_{\nlos} \backslash \{x\}} \Exp_{g_{y}} \big[ \exp \big( - s g_{y} \ell_{\nlos}(r_{y},h) \big) \big] \bigg] \\
\nonumber & = \Exp_{\Phi} \bigg[ \prod_{y \in \Phi_{\los} \backslash \{x\}} \frac{1}{1 + s \ell_{\los}(r_{y},h)} \\ 
\label{eq:L_I1} & \phantom{=} \ \times \prod_{y \in \Phi_{\nlos} \backslash \{x\}} \frac{1}{1 + s \ell_{\nlos}(r_{y},h)} \bigg] \\
\nonumber & = \exp \bigg( - 2 \pi \lambda \int_{\nu(r)}^{\infty} \bigg( 1 - p_{\los}(r) \frac{1}{1 + s \ell_{\los}(t,h)} \\
\label{eq:L_I2} & \phantom{=} \ - \big( 1 - p_{\los}(r) \big) \frac{1}{1 + s \ell_{\nlos}(t,h)} \bigg) t \diff t \bigg)
\end{align}
where in \eqref{eq:L_I1} we have applied the MGF of the exponential distribution and in \eqref{eq:L_I2} we have used the PGFL of a PPP. Finally, the expression in \eqref{eq:L_I} is obtained by including \eqref{eq:L_I_Q} with $\mathrm{Q} = \mathrm{NLOS}$ into \eqref{eq:L_I2}. \hfill \IEEEQED

%=========================================================================
\section{Effect of BS Height} \label{sec:A2}
%=========================================================================
\subsection{Proof of Lemma~\ref{lem:interference}} \label{sec:A2_interference}
%=========================================================================

Assume that the points of $\Phi$ are indexed such that their distances from the typical UE are in increasing order, i.e., $\{ r_{x_{i}} \leq r_{x_{i+1}} \}_{i=1}^{\infty}$. For strongest BS association, the expected interference power is given by

\begin{align}
\hspace{-2mm} \Exp \bigg[ \sum_{y \in \Phi \backslash \{x\}} \! g_{y} \ell(r_{y},h) \bigg] \! & < \Exp \bigg[ \sum_{x_{i} \in \Phi} g_{x_{i}} \ell(r_{x_{i}},h) \bigg] \\
\label{eq:interference1} & = \sum_{i=1}^{\infty} \Exp \big[ g_{x_{i}} \ell(r_{x_{i}},h) \big] \\
\label{eq:interference2} & = \sum_{i=1}^{\infty} \Exp \big[ \ell(r_{x_{i}},h) \big] \\
\label{eq:interference3} & = \sum_{i=1}^{\infty} \int_{0}^{\infty} \! (r^{2} + h^{2})^{-\frac{\alpha}{2}} f_{r_{x_{i}}}(r) \diff r
\end{align}
where \eqref{eq:interference2} follows from $\Exp \big[ g_{y} \ell(r_{y},h) \big] = \Exp [g_{y}] \Exp \big[ \ell(r_{y},h) \big]$ with $\Exp [g_{y}] = 1$ and where $f_{r_{x_{i}}}(r)$ in \eqref{eq:interference3} is the pdf of the distance between the typical UE and the $i$-th closest BS \cite[Ch.~2.9]{Hae12}: \vspace{-1mm}
\begin{align}
f_{r_{x_{i}}}(r) \triangleq e^{- \pi \lambda r^{2}} \frac{2 (\pi \lambda r^{2})^{i}}{r \Gamma(i)}.
\end{align}
Solving the integral in \eqref{eq:interference3} gives the expression in the right-hand side of \eqref{eq:interference}. On the other hand, the integral when $i=1$ yields the expected interference power from the closest interfering BS in \eqref{eq:interference_nearest}. Evidently, since the terms in the summation in \eqref{eq:interference1} are strictly decreasing with $i$ and the dominant interference term \eqref{eq:interference_nearest} is finite, then the aggregate interference power is also finite. \hfill \IEEEQED \vspace{-1mm}

%=========================================================================
\subsection{Proof of Theorem~\ref{th:height}} \label{sec:A2_height}
%=========================================================================

Consider the single-slope pathloss function $\ell(r_{x},h) = (r^{2} + h^{2})^{-\frac{\alpha}{2}}$. For closest BS association, we have
\begin{align}
\nonumber & \Pcov^{(\rmC)} (\theta, \lambda) \\
\label{eq:P_cov_C1} & = 2 \pi \lambda \! \int_{0}^{\infty} \! \! \exp \! \bigg( \! \! - 2 \pi \lambda \! \int_{r}^{\infty} \! \! \bigg( \! 1 - \frac{1}{1 + \theta \frac{\ell (t, h)}{\ell (r, h)}} \bigg) t \diff t \! \bigg) e^{- \pi \lambda r^{2}} \! r \diff r \\
& = 2 \pi \lambda \! \int_{0}^{\infty} \! \exp \big( - \pi \lambda \psi (\theta) (r^{2} + h^{2}) \big) e^{- \pi \lambda r^{2}} r \diff r \\
\label{eq:P_cov_C2} & = 2 \pi \lambda \! \int_{0}^{\infty} \! \exp \big( - \pi \lambda (\psi (\theta) + 1) r^{2} \big) r \diff r \exp \big( - \pi \lambda h^{2} \psi(\theta) \big)
\end{align}
where the inner integral in \eqref{eq:P_cov_C1} can be solved by substituting $\sqrt{t^{2} + h^{2}} \to t_{h}$ and with $\psi(z)$ given in \eqref{eq:psi}; finally, solving the integral in \eqref{eq:P_cov_C2} yields $\mathrm{P}_{\mathrm{cov},0}^{(\rmC)} (\theta)$ in \eqref{eq:P_cov_C_0}. On the other hand, for strongest BS association, we have
\begin{align}
\nonumber & \Pcov^{(\rmC)} (\theta, \lambda) \\
& = 2 \pi \lambda \! \int_{0}^{\infty} \! \! \exp \! \bigg( \! \! - 2 \pi \lambda \! \int_{0}^{\infty} \! \! \bigg( \! 1 - \frac{1}{1 + \theta \frac{\ell (t, h)}{\ell (r, h)}} \bigg) t \diff t \! \bigg) \! r \diff r \\
\label{eq:P_cov_S1} & = 2 \pi \lambda \! \int_{h}^{\infty} \! \! \exp \! \bigg( \! \! - 2 \pi \lambda \! \int_{h}^{\infty} \! \! \bigg( \! 1 - \frac{1}{1 + \theta r_{h}^{\alpha} t_{h}^{-\alpha}} \bigg) t_{h} \diff t_{h} \! \bigg) \! r_{h} \diff r_{h}
\end{align}
where in \eqref{eq:P_cov_S1} we have substituted $\sqrt{t^{2} + h^{2}} \to t_{h}$ in the inner integral and $\sqrt{r^{2} + h^{2}} \to r_{h}$ in the outer integral; finally, solving the inner integral in \eqref{eq:P_cov_S1} yields $\mathrm{P}_{\mathrm{cov},0}^{(\rmS)} (\theta)$ in \eqref{eq:P_cov_S}. \hfill \IEEEQED

%=========================================================================
\subsection{Proof of Lemma~\ref{lem:limits}} \label{sec:A2_limits}
%=========================================================================

First, (i) can be easily obtained from Theorem~\ref{th:height}. Furthermore, (ii) is a consequence of Lemma~\ref{lem:Laplace}. Lastly, (iii) follows from Theorem~\ref{th:height} for closest BS association and from
\begin{align}
\nonumber & \Pcov^{(\rmS)} (\theta, \lambda) \\
& = 2 \pi \lambda \! \int_{0}^{\infty} \! \! \exp \! \bigg( \! \! - 2 \pi \lambda \! \int_{0}^{\infty} \! \! \bigg( \! 1 - \frac{1}{1 + \theta \frac{\ell (t, h)}{\ell (r, h)}} \bigg) t \diff t \! \bigg) r \diff r \\
& < 2 \pi \lambda \! \int_{0}^{\infty} \! \! \exp \! \bigg( \! \! - 2 \pi \lambda \! \int_{r}^{\infty} \! \! \bigg( \! 1 - \frac{1}{1 + \theta \frac{\ell (t, h)}{\ell (r, h)}} \bigg) t \diff t \! \bigg) r \diff r \\
& = 2 \pi \lambda \int_{0}^{\infty} \! \! \exp \big( \! - \! \pi \lambda \psi (\theta) r^{2} \big) r \diff r \exp \big( \! - \! \pi \lambda h^{2} \psi(\theta) \big)
\end{align}
for strongest BS association (see Appendix~\ref{sec:A2_height} for details).~\IEEEQED

\vspace{2mm}

\addcontentsline{toc}{chapter}{References}
\bibliographystyle{IEEEtran}
\bibliography{IEEEabrv,ref_Huawei}

\begin{thebibliography}{10}
\providecommand{\url}[1]{#1}
\csname url@samestyle\endcsname
\providecommand{\newblock}{\relax}
\providecommand{\bibinfo}[2]{#2}
\providecommand{\BIBentrySTDinterwordspacing}{\spaceskip=0pt\relax}
\providecommand{\BIBentryALTinterwordstretchfactor}{4}
\providecommand{\BIBentryALTinterwordspacing}{\spaceskip=\fontdimen2\font plus
\BIBentryALTinterwordstretchfactor\fontdimen3\font minus
  \fontdimen4\font\relax}
\providecommand{\BIBforeignlanguage}[2]{{%
\expandafter\ifx\csname l@#1\endcsname\relax
\typeout{** WARNING: IEEEtran.bst: No hyphenation pattern has been}%
\typeout{** loaded for the language `#1'. Using the pattern for}%
\typeout{** the default language instead.}%
\else
\language=\csname l@#1\endcsname
\fi
#2}}
\providecommand{\BIBdecl}{\relax}
\BIBdecl

\bibitem{Bhu14}
N.~Bhushan, J.~Li, D.~Malladi, R.~Gilmore, D.~Brenner, A.~Damnjanovic,
  R.~Sukhavasi, C.~Patel, and S.~Geirhofer, ``Network densification: the
  dominant theme for wireless evolution into {5G},'' \emph{{IEEE} Commun.
  Mag.}, vol.~52, no.~2, pp. 82--89, Feb. 2014.

\bibitem{And11}
J.~G. Andrews, F.~Baccelli, and R.~K. Ganti, ``A tractable approach to coverage
  and rate in cellular networks,'' \emph{{IEEE} Trans. Commun.}, vol.~59,
  no.~11, pp. 3122--3134, Nov. 2011.

\bibitem{Dhi12}
H.~S. Dhillon, R.~K. Ganti, F.~Baccelli, and J.~G. Andrews, ``Modeling and
  analysis of {$K$}-tier downlink heterogeneous cellular networks,''
  \emph{{IEEE} J. Sel. Areas Commun.}, vol.~30, no.~3, pp. 550--560, Apr. 2012.

\bibitem{Zha15}
X.~Zhang and J.~G. Andrews, ``Downlink cellular network analysis with
  multi-slope path loss models,'' \emph{{IEEE} Trans. Wireless Commun.},
  vol.~63, no.~5, pp. 1881--1894, May 2015.

\bibitem{Gal15}
C.~Galiotto, N.~K. Pratas, N.~Marchetti, and L.~Doyle, ``A stochastic geometry
  framework for {LOS}--{NLOS} propagation in dense small cell networks,'' in
  \emph{Proc. {IEEE} Int. Conf. Commun. (ICC)}, London, UK, June 2015, pp.
  2851--2856.

\bibitem{Din16}
M.~Ding, P.~Wang, D.~Lopez-Perez, G.~Mao, and Z.~Lin, ``Performance impact of
  {LoS} and {NLoS} transmissions in dense cellular networks,'' \emph{{IEEE}
  Trans. Wireless Commun.}, vol.~15, no.~3, pp. 2365--2380, Mar. 2016.

\bibitem{Gup16}
A.~K. Gupta, M.~N. Kulkarni, E.~Visotsky, F.~W. Vook, A.~Ghosh, J.~G. Andrews,
  and R.~W. Heath, ``Rate analysis and feasibility of dynamic {TDD} in {5G}
  cellular systems,'' in \emph{Proc. {IEEE} Int. Conf. Commun. (ICC)}, May
  2016, pp. 1--6.

\bibitem{Arn16}
J.~Arnau, I.~Atzeni, and M.~Kountouris, ``Impact of {LOS/NLOS} propagation and
  path loss in ultra-dense cellular networks,'' in \emph{Proc. {IEEE} Int.
  Conf. Commun. (ICC)}, Kuala Lumpur, Malaysia, May 2016, pp. 1--6.

\bibitem{Atz17b}
\BIBentryALTinterwordspacing
I.~Atzeni, J.~Arnau, and M.~Kountouris, ``Downlink cellular network analysis
  with {LOS/NLOS} propagation and elevated base stations,'' Mar. 2017.
  [Online]. Available: \url{https://arxiv.org/pdf/1703.01279.pdf}
\BIBentrySTDinterwordspacing

\bibitem{And16}
J.~G. Andrews, T.~Bai, M.~N. Kulkarni, A.~Alkhateeb, A.~K. Gupta, and R.~W.
  Heath, ``Modeling and analyzing millimeter wave cellular systems,''
  \emph{{IEEE} Trans. Commun.}, vol.~65, no.~1, pp. 403--430, Jan. 2017.

\bibitem{Ngu16}
V.~M. Nguyen and M.~Kountouris, ``Coverage and capacity scaling laws in
  downlink ultra-dense cellular networks,'' in \emph{Proc. {IEEE} Int. Conf.
  Commun. (ICC)}, Kuala Lumpur, Malaysia, May 2016, pp. 1--6.

\bibitem{Bai14}
T.~Bai, R.~Vaze, and R.~W. Heath, ``Analysis of blockage effects on urban
  cellular networks,'' \emph{{IEEE} Trans. Wireless Commun.}, vol.~13, no.~9,
  pp. 5070--5083, Sep. 2014.

\bibitem{Din16a}
M.~Ding and D.~Lopez-Perez, ``Please lower small cell antenna heights in
  {5G},'' in \emph{Proc. {IEEE} Global Commun. Conf. Exhib. and Ind. Forum
  (GLOBECOM)}, Washington, DC, USA, Dec. 2016, pp. 1--6.

\bibitem{And16a}
J.~G. Andrews, X.~Zhang, G.~D. Durgin, and A.~K. Gupta, ``Are we approaching
  the fundamental limits of wireless network densification?'' \emph{{IEEE}
  Commun. Mag.}, vol.~54, no.~10, pp. 184--190, Oct. 2016.

\bibitem{Lee16}
J.~Lee, X.~Zhang, and F.~Baccelli, ``A {3-D} spatial model for in-building
  wireless networks with correlated shadowing,'' \emph{{IEEE} Trans. Wireless
  Commun.}, vol.~15, no.~11, pp. 7778--7793, Nov. 2016.

\bibitem{Hae12}
M.~Haenggi, \emph{Stochastic Geometry for Wireless Networks}.\hskip 1em plus
  0.5em minus 0.4em\relax New York, NY, USA: Cambridge University Press, 2012.

\end{thebibliography}

\end{document}